\documentclass[twocolumn,aps,amsmath,amssymb]{revtex4}

\usepackage{hyperref}

\newcommand{\Ga}{\Gamma}

\newcommand{\cI}{{\cal I}}

\newcommand{\Ba}{{\bar a}}
\newcommand{\Bb}{{\bar b}}
\newcommand{\Bc}{{\bar c}}

\newcommand{\bA}{{\bf 8}}

\newcommand{\ha}{{\hat a}}
\newcommand{\hb}{{\hat b}}

\newcommand{\cJ}{{\mathcal J}}
\newcommand{\rX}{{\rm X}}
\newcommand{\rE}{{\rm E}}
\newcommand{\rA}{{\rm A}}
\newcommand{\ta}{{\tt{a}}}
\newcommand{\tb}{{\tt{b}}}
\newcommand{\tc}{{\tt{c}}}
\newcommand{\td}{{\tt{d}}}

\newcommand{\kof}[1]{{\rm k}_{#1}}

\newcommand{\veps}{\varepsilon}
\newcommand{\Gt}{\tilde\Gamma}

\newcommand{\beq}{\begin{equation}}
\newcommand{\eeq}{\end{equation}}
\newcommand{\bea}{\begin{eqnarray}}
\newcommand{\eea}{\end{eqnarray}}
\newcommand{\nn}{\nonumber}

\newcommand{\ra}{\rightarrow}
\newcommand{\E}{{\rm E}_{10}}
\newcommand{\KE}{K({\rm E}_{10})}

\begin{document}

\title{Standard Model Fermions and $\KE$}
\author{Axel Kleinschmidt$^{1,2}$ and Hermann Nicolai$^1$}
\address{$^1$ Max-Planck-Institut f\"ur Gravitationsphysik
(Albert-Einstein-Institut)\\
M\"uhlenberg 1, DE-14476 Potsdam, Germany\\
$^2$ International Solvay Institutes, ULB-Campus Plaine, CP 231, BE-1050 Bruxelles, Belgium}

\vspace{5mm}

\begin{abstract} 
In recent work \cite{MN} it was shown how to rectify Gell-Mann's proposal for identifying
the 48 quarks and leptons of the Standard Model with the 48 spin-$\frac12$ fermions of 
maximal SO(8) gauged supergravity remaining after the removal of eight  Goldstinos,
by deforming the residual U(1) symmetry at the SU(3)$\,\times\,$U(1) stationary point 
of $N=8$ supergravity, so as to also achieve agreement of the electric  charge assignments. 
In this Letter we show that the required deformation, while not in SU(8), does belong 
to $\KE$, the `maximal compact' subgroup of $\E$ which is a possible candidate symmetry 
underlying M theory. The incorporation of infinite-dimensional Kac--Moody symmetries 
of hyperbolic type, apparently unavoidable for the present scheme to work,
 opens up completely new perspectives on embedding Standard Model physics into 
a Planck scale theory of quantum gravity.
\end{abstract}

\maketitle 

The question whether or not the maximally extended $N=8$ supergravity theory \cite{CJ,dWN}
can be related to Standard Model physics has been under debate for a long time. 
Very recent work \cite{MN} has taken up an old proposal of Gell-Mann's  \cite{GellMann} 
on how to match the 48 quarks and leptons (including right-chiral neutrinos) of the 
Standard Model with the 48 spin-$\frac12$ fermions of maximal SO(8) gauged supergravity  
that remain after the removal of eight Goldstinos (as required by the complete breaking 
of $N=8$ supersymmetry). This scheme, which was subsequently shown to be realised at the 
SU(3)$\,\times\,$U(1) stationary point of maximal gauged SO(8) supergravity \cite{NW}, 
relies on identifying the residual SU(3) of supergravity with the diagonal subgroup 
of the colour group SU(3)$_c$ and a new family symmetry SU(3)$_f$. Intriguingly, in this
way complete agreement is found in the SU(3) charge assignments of quarks and leptons
and the spin-$\frac12$ fermions of $N=8$ supergravity, but there remained a 
systematic mismatch in the electric charges by a spurion charge of $q=\pm \frac16$. 
The main advance reported in \cite{MN} was to identify the `missing'  U(1)$_q$ that 
rectifies this mismatch, and that  was found to take a surprisingly simple form. However,
this deformation cannot be explained from `within' $N=8$ supergravity  
(nor from a hypothetical embedding of maximal  gauged supergravity 
into the known superstring theories), as U(1)$_q$ is not contained in its 
R symmetry group SU(8).  In this Letter we show that the required deformation is, 
however, contained in an infinite dimensional extension of SU(8), namely the 
involutory  `maximal compact' subgroup $\KE$ of the hyperbolic Kac--Moody group 
$\E$, which has been proposed as a possible candidate symmetry of M theory 
\cite{DHN}.\footnote{For an earlier and conceptually different proposal based on the non-hyperbolic Kac--Moody algebra $\rE_{11}$ see~\cite{West}.} This, we believe, places the question stated above, and also
the eventual incorporation of the chiral electroweak gauge interactions
(not considered in \cite{GellMann,NW}), in an entirely new context, by embedding 
(at least a subset of) the Standard Model symmetries into an infinite-dimensional 
extension of the exceptional duality symmetries of maximal supergravity. 
This approach, never tried before to the best of our knowledge, offers completely new 
perspectives on the possible Planck scale origin of Standard Model physics.

For the rest of this text we will concentrate on the fermionic sector of $N\!=\!8$ supergravity, 
which consists of eight gravitinos $\psi_\mu^i$ transforming in the $\bA$, and a tri-spinor 
of spin-$\frac12$ fermions $\chi^{ijk}$  transforming in the $\bf{56}$ of SU(8), whence  
$\chi^{ijk}$ is fully antisymmetric in the SU(8) indices $i,j,k$, with (positive and negative)
chirality corresponding to (upper and lower) position of the indices, and $\chi^{ijk} =
(\chi_{ijk})^*$. Here we will, however, restrict attention to the vector-like 
SO(8) subgroup of SU(8), for which the distinction between upper and lower indices 
is immaterial, whence we will not distinguish between $\chi^{ijk}$ and 
$\chi_{ijk}$ in the remainder. The residual vector-like SO(8) transformations act as
\beq\label{Uchi}
\chi^{ijk} \; \ra \; U^i{}_l U^j{}_m U^k{}_n \chi^{lmn} \qquad
\mbox{with $U \in$ SO(8)}.
\eeq

In order to obtain the correct electric charge assignments of the quarks and leptons
it was found in \cite{MN} that the U(1) subgroup of SU(3)$\,\times\,$U(1) must be deformed 
by a new (still vector-like) U(1)$_q$ whose action on the tri-spinor $\chi^{ijk}$ is generated 
by the following 56-by-56 matrix 
\beq\label{cI}
\cI := \frac12 \Big(T \wedge {\bf 1} \wedge {\bf 1} \, + \,  {\bf 1} \wedge T \wedge {\bf 1} \, + \,
        {\bf 1} \wedge {\bf 1} \wedge T \, + \,  T \wedge T \wedge T \Big)
\eeq
acting in the $\bA\wedge\bA\wedge\bA$ representation of SO(8). Here 
 \begin{equation} \label{T}
 T \,=\, 
               \begin{pmatrix}
                0 & 1  & 0 & 0 & 0 & 0 & 0 & 0 \\
                -1  & 0 & 0 & 0 & 0 & 0 & 0 & 0 \\
                0 & 0 & 0 & 1 & 0 & 0 & 0 & 0 \\
                0 & 0 & -1 & 0 & 0 & 0 & 0 & 0 \\
                0 & 0 & 0 & 0 & 0 &  1 & 0 & 0 \\
                0 & 0 & 0 & 0 & -1 & 0 & 0 & 0 \\
                0 & 0 & 0 & 0 & 0 & 0 & 0 & 1\\
                0 & 0 & 0 & 0 & 0 & 0 & -1 & 0
                \end{pmatrix},
\end{equation}
represents the imaginary unit in the breaking of SO(8) to SU(3)$\,\times\,$U(1).
We note that, from $T^2 = -{\bf 1}$ we have $\cI^2 \, = \, - {\bf 1}$, whence 
(\ref{cI}) can be trivially exponentiated to a U(1)$_q$ phase rotation. 
The combination (\ref{cI}) differs from the usual co-product obtained from (\ref{Uchi}) 
with $U = \exp (\omega T)$ by the 56-by-56 matrix $T\wedge T \wedge T$. Importantly,
the latter is not in SU(8), although it does commute with
the SU(3)$\,\times\,$U(1) subgroup of SO(8), and hence merely {\em deforms} this 
subgroup, but does not enlarge it.  We will now show how to accommodate the 
triple wedge product $T\wedge T \wedge T$ by enlarging the R symmetry SU(8) 
of $N\!=\!8$ supergravity to the bigger, and in fact, infinite-dimensional R symmetry 
$\KE$, in accordance with the anticipated enlargement of the finite-dimensional
exceptional dualities of maximal supergravities to infinite-dimensional groups. 

To proceed we recall how the fermions of $D\!=\! 11$ supergravity \cite{CJS}
are related to those of $N\!=\!8$ supergravity \cite{CJ,dWN1}. Denoting the (spatial) 
$D=11$ gravitino  components by $\Psi^a_A$ (with $a,b,...= 1,...,10$ and $D=11$ spinor 
indices $A,B,...=1,...,32$) and adopting the temporal supersymmetry gauge $\Psi^0_A = (\Gamma^0\Gamma_a)_{AB} \Psi^a_B$ as in \cite{DKN1,DKN}, 
we split the $D=11$ gravitino into four-dimensional spatial and internal components as follows
\beq
\Psi^a_A = \big(\Psi^\ha_{\alpha i} \;, \, \Psi^\Ba_{\alpha i} \big)
\eeq
with flat spatial indices $\ha,\hb,...= 1,2,3$ and flat internal indices $\Ba, \Bb,...= 4,\dots,10$, 
whose position again does not matter as they are pulled up and down with
$\delta_{ab}$. The $D=11$ spinor indices $A,B,...$ are split as $A\equiv (\alpha,i)$
into $D=4$ spinor indices 
$\alpha,\beta,...= 1,...,4$  and internal SO(8) indices $i,j,...=1,...,8$ (whose position likewise
does not matter here as we restrict attention to a vector-like symmetry). Ignoring a Weyl rescaling 
factor and a chiral redefinition, and not making a split into left-chiral and right-chiral
components as in  \cite{dWN1},  we have
\bea\label{psichi}
\psi^i_{\ha \alpha} &\;\propto\;& \Psi^i_{\ha \alpha}   - 
\frac12  \sum_{\Bc=4}^{10} \Gamma^\Bc_{ij} 
\Big(\gamma^5 \gamma_\ha \Psi^j_\Bc\Big)_\alpha     \\ \label{psichi1}
\chi^{ijk}_{ \alpha} &\;\propto\;& \sum_{\Ba=4}^{10} \Gamma^\Ba_{[ij} \Psi^\Ba_{k]\alpha}
\eea 
where we temporarily suspend the summation convention for the indices
$a,b,...$ (the summation convention remains, however, in force for all other indices).
For the  implementation of the action of $T\wedge T \wedge T$ we also need
the following redefinition of the $D=11$ gravitino \cite{DH}:
\beq\label{DH} 
\Phi^\ta_A = \Ga^a_{AB} \Psi^a_B
\qquad \mbox{(no sum on $a$!)}
\eeq
Because there is {\em no} summation on the spatial index $a$, manifest 
SO(10) covariance is lost. To emphasise this point we adopt a different font 
($\ta,\tb,...$) although these indices have the same range as $a,b,....$ before \cite{KN}.
Importantly, however, the position of the indices $\ta,\tb,...$ now {\em does} matter, as 
they are to be raised and lowered with the (Lorentzian) DeWitt metric and its inverse 
\beq\label{Gab}
G_{\ta\tb} = \delta_{\ta\tb} - 1 \quad \Leftrightarrow\quad
G^{\ta\tb} = \delta^{\ta\tb} - \frac19.
\eeq
With the redefinition (\ref{DH}) the formula (\ref{psichi1}) becomes
\beq
\chi_{ijk \, \alpha} \;\propto \; 
\sum_{\ta = 4}^{10} \Gamma^\ta_{[ij} \Gamma^\ta_{kl]} \Phi^\ta_{l \alpha}.
\eeq
The action of $T\wedge T \wedge T$ is therefore realised via (now suppressing $D=4$ 
spinor indices)
\bea\label{Tchi}
\chi_{ijk} \; &\rightarrow& \;\; T_{il} \,T_{jm} \, T_{kn}  \, \chi_{lmn}    \nn\\[2mm]
 &\propto &  \, \sum_{\ta=4}^{10} \, (T\Gamma^\ta T)_{[ij} (T \Gamma^\ta T)_{kl]} \, T_{lm}   \Phi_m^{\ta }   
\eea
where we have inserted a factor $TT= - {\bf 1}$ and used the antisymmetry of $T$. 
Next we recall that there is a representation of the SO(7) $\Gamma$-matrices where 
\beq\label{Gamma45}
T_{ij} \,=\, \Gamma^{45}_{ij}
\eeq 
(see e.g. appendix  E of \cite{GGKNP}); it is then easy to see that
\beq 
(T\Gamma^\ta T)_{[ij} (T \Gamma^\ta T)_{kl]}  = 
\Gamma^\ta_{[ij} \Gamma^\ta_{kl]}     
\eeq
even without summation over $\ta$. Using this formula we conclude from (\ref{Tchi})
that the desired action takes a very simple form on the redefined spinors (\ref{DH}), to wit,
\beq\label{TTT}
\Phi^\ta_{i \alpha} \;\;  \longrightarrow \quad T_{ij} \Phi^\ta_{j\alpha}
\eeq
which leaves the $D=4$ spinor indices unaffected. Of course, one could also
(though less elegantly) express this action in terms of the original spinors $\Psi^a_A$.
We stress that in order to preserve the relation (\ref{psichi}), (\ref{TTT}) must 
hold {\em for all} $\,\ta=1,...,10$. From this follows the action of the new generator
on the $D=4$ gravitino, an insight that the arguments in \cite{MN} could not provide. 
Observe that the redefinition $\psi_\ha^i \rightarrow \gamma^\ha \psi_\ha^i$ implied 
by (\ref{DH}) does not affect this conclusion, as $\gamma^\ha$ commutes with $T$.

We now want to show that the action (\ref{TTT}) is contained in $\KE$, the supposed 
R symmetry of M theory. We refer to our previous work \cite{DKN1,DKN,KN,KN1} for 
detailed explanations on $\KE$, and here simply summarise some salient results 
(see also \cite{dBHP,dBHP1} for 
related work). The group $\KE$ is the involutory subgroup of $\E$ which is left invariant
by the Cartan-Chevalley involution defined on $\E$ in terms of its Chevalley-Serre
presentation. As such, it contains the R symmetries of all $D \geq 2$ maximal 
supergravities as subgroups (and thus also {\em chiral} transformations for
even $D$); more specifically, we have
\beq
{\rm SU}(8)  \subset {\rm SO}(16) \subset K(\rE_9) \subset \KE
\eeq
The fermions transform in spinorial (double-valued) representations of $\KE$. A remarkable 
property of the algebra $\KE$ is that, though infinite-dimensional, it admits {\em finite-dimensional}, 
hence {\em unfaithful} representations \cite{DKN,dBHP}. These are the Dirac \cite{dBHP1}
and vector-spinor representations \cite{DKN,dBHP}, which can be directly deduced from 
$D=11$ supergravity (in addition, two `higher spin' realisations are known \cite{KN}).
As a consequence, $\KE$ is {\em not} simple, because it has nontrivial (finite codimension)
ideals $\cJ$ which are associated with the unfaithful representations in the way explained 
in \cite{DKN}. Accordingly, the quotient $\KE/\cJ$ is a finite-dimensional group; 
more specifically,  denoting the vector-spinor ideal by $\cJ_{\rm vs}$, evidence was 
presented in \cite{KN} that
$
\KE \big/ \cJ_{\rm vs} \,= \,  {\rm Spin} (288,32).
$
The fact that the `compact' subgroup $\KE\subset\E$ in this way gives rise to a 
{\em non-compact} quotient group is another unusual feature of $\KE$.

A convenient realization of the $\KE$ Lie algebra generators in the vector-spinor representation 
was found in \cite{KN,KN1} (following earlier work on $K(\rA\rE_3)$ in \cite{DH,DS,Damour:2014cba}). Like
the generators of $\E$, the generators $\kof{\alpha}^r$ of $\KE$ can be labeled by $\E$ roots 
$\alpha$ and the associated multiplicity index $r$, but such that \cite{KN1}
\beq
\kof{\alpha}^r = - \kof{-\alpha}^r  \; , \quad \textrm{for all $\E$ roots $\alpha$.}
\eeq
As shown in \cite{KN}, for the vector spinor representation there 
is a concrete realization of these generators in terms of 320-by-320 matrices. For all
{\em real} roots $\alpha$ of $\E$ (for which the multiplicity label $r$ is not needed) we have
\beq\label{Jreal}
(\kof{\alpha})_{\ta A ,  \tb B} = \frac12 \rX_{\ta\tb}(\alpha) \Gt(\alpha)_{AB}
\eeq
where the symmetric matrix $\rX_{\ta\tb}$ is given by
\beq\label{rX}
\rX_{\ta\tb}(\alpha) = -\frac12 \alpha_\ta\alpha_\tb +\frac14 G_{\ta\tb}
\eeq 
in terms of the root components $\alpha_\ta$ in the `wall basis' used in \cite{KN};
indices $\ta,\tb$ are raised and lowered by means of (\ref{Gab}). As explained in \cite{KN}
there is a map from the $\E$ root lattice into the SO(10) Clifford algebra that associates
to each root $\alpha$ of $\E$ a particular element $\Gt(\alpha)= - \Gt(-\alpha)$ of the
Clifford algebra; furthermore the matrices $\Gt(\alpha)$ are anti-symmetric for 
$\alpha^2 \in 4 {\mathbb{Z}} +2$ and symmetric for $\alpha^2\in 4{\mathbb{Z}}$.
Because the SO(10) Clifford algebra is finite-dimensional, and because there 
are infinitely many real and imaginary roots of $\E$, it follows that infinitely many 
$\E$ roots $\alpha$ are mapped to the same element of the Clifford algebra.

To prove that (\ref{Jreal}) indeed generates the algebra $\KE$, one substitutes the 
ten simple roots of $\E$ into (\ref{Jreal})  and verifies the defining relations for $\KE$ \cite{KN} 
(the latter characterise the involutory subalgebra in a manner analogous to the 
Chevalley--Serre presentation for general Kac--Moody algebras  \cite{Berman,Kac}). 
The Lie algebra $\KE$ in the vector spinor representation is thus generated by taking 
commutators of the above real root generators in all possible ways. In this way 
one `reaches' all imaginary root spaces with $\alpha^2\leq 0$. However, due
to the unfaithfulness of the representation the image of the root space elements
consists of linear combinations of finitely many basis elements. The generating 
elements, and thus $\KE$, leave invariant the Lorentzian bilinear form 
\beq\label{Bilinear}
(V,W) \equiv G_{\ta\tb} V^\ta_A W^\tb_A\quad\textrm{(of signature $(288,32)$).}
\eeq

For general imaginary roots the formula (\ref{Jreal}) is no longer valid with (\ref{rX}). 
What is clear, however, is that all matrices $\kof{\alpha}^r$ generated in this way are 
antisymmetric under interchange of the index pairs $(\ta A)$ and $(\tb B)$, that is,
\beq
(\kof{\alpha}^r)_{\ta A , \tb B} = - (\kof{\alpha}^r)_{\tb B , \ta A}. 
\eeq
and can thus be written as a linear combination of matrices of the form (\ref{Jreal}), 
with either $\rX_{\ta\tb}$ symmetric in $(\ta\tb)$ and $\Gt(\alpha)_{AB}$ anti-symmetric in $[AB]$,
or antisymmetric in $[\ta\tb]$ and symmetric in $(AB)$. Because all such matrices
leave invariant the Lorentzian bilinear form (\ref{Bilinear}) they all belong to the Lie algebra 
of $\mathfrak{so}(288,32)$ \cite{KN}. 

Although we do not have a general formula for arbitrary imaginary roots, explicit formulas 
do exist for null roots $\delta$, and for certain time-like roots $\Lambda$ \cite{KN1}. 
For null roots $\delta$, we have
\beq\label{null}
(\kof{\delta}^r)_{\ta A, \tb B}  = \veps^r_{[\ta} \delta_{\tb ]}^{\ } \Gt(\delta)_{AB}
\eeq
with eight transversal polarisation vectors $\veps^r$. For time-like roots $\Lambda$ with
$\Lambda^2= 2 - 4n$ (for $n\geq 0$), the corresponding $\kof{\Lambda}^r$ can be realised in the 
form  (\ref{Jreal}) by choosing a decomposition $\Lambda= \alpha + \beta$ with $\alpha^2 =
\beta^2 = 2$ and $\alpha\cdot\beta = -(2n+1)$; this gives
\beq\label{Lab}
\rX_{\ta\tb}^{(\alpha)}(\Lambda) = 
- \frac12 \alpha_\ta \alpha_\tb - \frac12 \beta_\ta \beta_\tb
   - (2n+1) \alpha_{(\ta} \beta_{\tb)}  + \frac14 G_{\ta\tb}
\eeq
Taking $n=1$ (that is, $\Lambda^2=-2$) as an example and letting the decomposition range 
over all pairs of real roots $(\alpha,\beta)$ with $\Lambda =  \alpha + \beta$ one thus 
re-constructs the full root space, of dimension ${\rm mult}(\Lambda) =44$. For larger $n$
the multiplicity of $\Lambda$ increases rapidly\footnote{For instance, mult(2$\Lambda$) = 2\,472,
mult(3$\Lambda$)= 425\,058 and mult(4$\Lambda$)= 130\,593\,068.}, and one can no longer exhaust the full root space with the $\rX_{\ta\tb}^{(\alpha)}(\Lambda)$.

Returning to our initial problem we note that 
\beq
k_{\ta A, \tb B} = G_{\ta\tb} T_{AB} \equiv G_{\ta\tb} \delta_{\alpha\beta} T_{ij} \;
     \in \; \mathfrak{so}(288,32)
\eeq
whence this matrix can be generated by a linear combination of matrices obtained 
by multiple commutation of the basic $\KE$ generators (because a linear combination
may be required, we omit the root and multiplicity labels on $k$). 
To see how one can arrive at the requisite linear combination
we note that there are infinitely many roots $\alpha$ (both real and imaginary) that satisfy $\tilde{\Gamma}(\alpha) = T = \Gamma^{45}$.
 The task of finding a $\KE$ generator that implements $T\wedge T \wedge T$ of (\ref{TTT}) is then reduced to finding a combination of tensors $\rX_{\ta\tb}$ that equals $G_{\ta\tb}$. We are not aware of a single root that achieves this but establishing the existence of a linear combination can be achieved as follows. In accordance with (\ref{Lab}) one considers
the set of all $\rX_{\ta\tb}$ that can arise from the commutation of two real root 
generators $\rX_{\ta\tb}(\alpha)$ and $\rX_{\tc\td}(\beta)$ (given as in (\ref{rX})) such that $\alpha+\beta=\Lambda$ is an imaginary root that satisfies $\tilde{\Gamma}(\Lambda) = \Gamma^{45}$. Similarly, one can perform the same analysis for odd multiples of $\Lambda$ given by $(2k+1)\Lambda$ since then $\tilde{\Gamma}\big((2k+1)\Lambda\big)=\tilde{\Gamma}(\Lambda)=\Gamma^{45}$. We have shown by an explicit computer analysis that one can find a linear combination of the generated $\rX_{\ta\tb}\big((2k+1)\Lambda\big)$ that equals $G_{\ta\tb}$ and therefore the desired realization of $T\wedge T \wedge T$ on the spinors of $D=11$ supergravity within $\KE$. The generator just constructed only extends the R symmetry ${\rm SU}(8)\subset {\rm SO}(3)\times {\rm SU}(8)\subset \KE$ and thus leaves the spatial rotation ${\rm SO}(3)$ symmetry untouched.

The above argument demonstrates the existence of an element of $\KE$ that acts according to (\ref{TTT}), but the combination identified above does not necessarily have a simple algebraic interpretation. 
Because the spinors $\phi^{\ta}_A$ form an unfaithful representation of $\KE$ there are infinitely 
many elements that act in this way, and it is thus possible that an alternative realization of 
$T\wedge T \wedge T$ exists that has a simple physical origin. For the realisation
found here one already has to go up to level $\ell=18$ in a level decomposition of $\KE$
(that follows directly from the corresponding tables for $\E$ given in \cite{FN}); there is thus
no easy way of reproducing this result by simple iteration of the low level 
$\KE$ transformation rules given in \cite{DKN}.
The explicit realization of the charge shifting $U(1)_q$ generator 
above relies on the existence of time-like imaginary roots and their  integer 
multiples, but there may be other possibilities, in particular, using only real roots.
In any case, it does  not appear possible to construct the requisite element without  use of 
the `hyperbolic'  over-extended root of $\E$, since the structure of the root system of the affine 
subalgebra $\mathfrak{e}_9$ is too restricted. 
In this sense, the extension to the full {\em hyperbolic} Kac--Moody algebra and its
involutory subalgebra could be essential for linking $N\!=\!8$ supergravity to the
real world.

We note that the embedding of $T\wedge T \wedge T$ into $\KE$ in principle 
also allows for a realisation of this transformation on the bosonic fields of the
spinning $\E/\KE$ model studied in \cite{DKN}, although the ambiguities related to 
the unfaithfulness of the fermionic realisation of $\KE$ remain to be resolved.
More precisely, while there are infinitely many combinations of $\KE$ generators
that act in the same way on the fermions, these will act differently on the bosonic coset variables on which $\KE$ is realised faithfully. The bosonic variables can thus in principle be used to remove all ambiguities.

The results of \cite{MN} and this Letter represent a significant shift away from the standard paradigm of how to understand the possible emergence of the Standard Model fermions from 
a Planck scale unified theory, as for instance embodied in currently popular
superstring inspired scenarios of low energy ($N\!=\!1$) supergravity. There one starts
from a finite-dimensional {\em compact} Yang-Mills gauge group (such as $\rE_6 \times \rE_8$), 
with the fermions transforming  in a standard representation. This symmetry is  
assumed to be present {\em as a space-time-based symmetry} already
at the Planck scale, and then assumed to be broken in a cascade of symmetry 
reductions as one descends to the electroweak scale. By contrast, the present 
scheme proceeds from an infinite-dimensional group that can be fully present as 
a symmetry only in a phase of the theory {\em prior to the emergence of classical 
space and time}, in accord with the proposal of \cite{DHN}, and {\em crucially relies on the 
infinite-dimensionality of this group} (and the associated Kac--Moody algebra).\footnote{A curious observation also concerns the discrete subgroups of $\KE$: The finite group ${\rm PSL}_2(7)$, a maximal discrete 
subgroup of the family symmetry ${\rm SU}(3)_f$ which was recently invoked to explain
the quark mass hierarchy~\cite{Ramond}
sits naturally in the Weyl group of ${\rm E}_7$. The latter is contained in 
the Weyl group of ${\rm E}_{10}$, and thus also inside $\KE$.}
We emphasise once again that $\KE$ does possess 
chirality, offering new perspectives for the incorporation of chiral gauge symmetries, 
such that the electroweak sector of  the Standard Model may eventually be understood 
in a way very different from currently prevailing views.

\vspace{5mm}

\noindent
 {\bf Acknowledgments:}  HN thanks the University of Western Australia, and especially
 Ian McArthur and Sergei Kuzenko for their hospitality and support during a most enjoyable 
 stay while this work was being carried out.
 The authors are also grateful to Thibault Damour for useful discussions.


\begin{thebibliography}{99}

\bibitem{MN} K.A.~Meissner and H.~Nicolai, 
Phys. Rev. {\bf D 91} (2015) 065029 

\bibitem{CJ} E.~Cremmer and B.~Julia, Nucl. Phys. {\bf B159} (1979) 141.

\bibitem{dWN} B.~de Wit and H.~Nicolai, Nucl. Phys. {\bf B208} (1982) 323.

\bibitem{GellMann} M.~Gell-Mann, in Proceedings of the Shelter Island Meeting II (1983),
       Caltech Preprint CALT-68-1153.

\bibitem{NW} H.~Nicolai and N.P.~Warner, Nucl. Phys. {\bf B259} (1985) 412.

\bibitem{DHN}
  T.~Damour, M.~Henneaux and H.~Nicolai,
  Phys.\ Rev.\ Lett.\  {\bf 89} (2002) 221601.
  
\bibitem{West} 
 P.C.~West, 
 Class.Quant.Grav. {\bf 18} (2001) 4443.


\bibitem{CJS}
  E.~Cremmer, B.~Julia and J.~Scherk,
  Phys.\ Lett. {\bf B76} (1978) 409.

\bibitem{dWN1} B.~de Wit and H.~Nicolai, Nucl. Phys. {\bf B274} (1986) 36.
  
\bibitem{DKN1}
  T.~Damour, A.~Kleinschmidt and H.~Nicolai,
  Phys.\ Lett.\ {\bf B 634} (2006) 319.

\bibitem{DKN}
  T.~Damour, A.~Kleinschmidt and H.~Nicolai,
  JHEP {\bf 0608} (2006) 046.

\bibitem{DH}
  T.~Damour and C.~Hillmann,
  JHEP {\bf 0908} (2009) 100.

\bibitem{KN}
  A.~Kleinschmidt and H.~Nicolai,
  JHEP {\bf 1308} (2013) 041.

\bibitem{KN1} A.~Kleinschmidt, H.~Nicolai and N.K.~Chidambaram,  Phys. Rev.  {\bf D91} (2015) 085039.
           

\bibitem{GGKNP} H.~Godazgar, M.~Godazgar, O.~Kr\"uger, H.~Nicolai and K.~Pilch,
 JHEP {\bf 1501} (2015) 056.

\bibitem{dBHP}
  S.~de Buyl, M.~Henneaux and L.~Paulot,
  JHEP {\bf 0602} (2006) 056.

\bibitem{dBHP1}
  S.~de Buyl, M.~Henneaux and L.~Paulot,
  Class.\ Quant.\ Grav.\  {\bf 22} (2005) 3595.

\bibitem{DS}
  T.~Damour and P.~Spindel,
  Class.\ Quant.\ Grav.\  {\bf 30} (2013) 162001.
  
\bibitem{Damour:2014cba}
  T.~Damour and P.~Spindel,
  Phys.\ Rev.\ D {\bf 90} (2014) 10,  103509.
  
\bibitem{Berman} S.~Berman, Commun. Algebra {\bf 17} (1989) 3165.

\bibitem{Kac} V.~Kac, {\em Infinite dimensional Lie Algebras}, 3rd edition,
    Cambridge University Press (1990).

\bibitem{FN} H.~Nicolai and T.~Fischbacher, 
{\tt hep-th/0301017}.


\bibitem{Ramond} 
G.~Chen, M.~J.~P\'erez and P.~Ramond,
{\tt arXiv:1412.6107}



\end{thebibliography}
\end{document}